# Service Discovery in Mobile Ad hoc Networks Using Association Rules Mining

Noman Islam, Zubair A. Shaikh
Center of Research in Ubiquitous Computing
National University of Computer & Emerging Sciences
Karachi, Pakistan

*Abstract*- **Service Discovery in Mobile Ad hoc Networks (MANET) is a difficult task because of the dynamic nature of such networks. The availability of the services in MANET varies with respect to time and location. However, the service discovery requests issued by a consumer in one session are correlated, in most of the cases. If we can find out the correlation among the services and piggyback one service response with others in advance, we can gain improvement in performance. By utilizing this correlation property among service discovery requests, we have proposed a novel approach to Service Discovery in Mobile Ad hoc Networks. We have simulated the proposed approach in JIST/SWANS simulator and the results have shown significant performance improvement.**

## I. INTRODUCTION

Mobile Ad hoc Networks are infrastructure less networks that are formed among a number of mobile nodes[1]. Due to the dynamic nature of these types of networks, algorithms that were designed for traditional networks (e.g. routing, transport protocols, security and QoS) don't show good performance for MANET. As a result active research work is going on to propose new algorithms and protocols for MANET. Service Discovery in MANET is one such issue that has gained considerable attention during recent years. A service is any tangible or non-tangible resource that one can invoke to gain some utility. Service Discovery can be defined as the process of discovering some service on behalf of its user based on the preferences (e.g. QoS) specified by the user and constraints (e.g. cost) specified by the provider.

Service Discovery in MANET is a very challenging task due to a number of issues. One of the issues is to decide where to place the directory or list of the services available on the network. Because there is no prior infrastructure in MANET, we can't dedicate any node or set of nodes as directory server, in advance. Usually, the directory is distributed among the nodes in MANET. Another important issue is the actual discovery of the service in the network. If there are few dedicated nodes, they can be queried for the service. Otherwise, broadcasting or their variations could be used for service discovery. Since the nodes are mobile and the changes (like link failure, node failure etc.) are quite common in such networks, maintaining a consistent view of the service availability in such networks is also an important task.

Service Discovery in MANET provides a number of useful benefits to the end user. For example: if you are traveling on a car and assuming an Ad hoc network is formed between vehicles around you. You can utilize a number of location based services while roaming along the road. This includes road safety services (e.g. weather warning, accident information, alternate route assistance) or non safety services (e.g. M-Banking, locating a restaurant/parking/fuel station). In most of these cases, a number of service discovery requests are issued simultaneously. We can call such simultaneous requests as belonging to one **service session**. Naturally, the service discovery requests that are issued in one session are correlated. For example, if the user intention is shopping, his requests will be to find out the shopping malls around it, the shopping items available at the mall, the cost of individual items etc. Our motivation is that if we can some how find out correlation among the services of one session, we can predict future discovery requests based on current service discovery request. In our research, we have used the Association Rules mining algorithm to achieve this objective. Using the correlation among the services and piggybacking future service request answers along with current service requests, we have proved significant gain in performance. Rest of the sections of this paper has been organized as follows: First we will review the literature on service discovery requests. Then we will discuss the proposed approach and its implementation details along with results. Finally, we will conclude the paper with discussions on future work.

## II. LITERATURE REVIEW

Work on Service Discovery in Ad hoc Networks have been reported from various perspectives in literature. Most of the classical algorithms for Service Discovery can be classified as directory-less and directory-based approaches. Among the first type, we have Allia[2], Konark[3], GSD[4], IBM DEAPSpace [5] and Microsoft UPnP [6] algorithm. The basic concept in **Allia** and **GSD** is the formation of alliance and groups respectively to reduce service discovery overhead. **Konark** represents the services in WSDL format and is based on broadcasting of service advertisement and then caching of these advertisements by peer nodes. **IBM DEAPSpace algorithm**, targeted for short range devices is based on intelligent broadcasting of services information to single hop devices. **UPnP** is a P2P protocol that uses broadcasting for service discovery and is based on SOAP for service invocation. Among the directory based approaches we have **JINI**[7], **DReggie**[8] and Mallah[9]. **JINI** is proposed for distributed environment, is based on Java RMI mechanism and uses Java Interfaces as service identifiers. **DReggie** is an extension of JINI that adds semantics to it by representing services through an ontology language called DAML. **SLP[10]** proposed by IETF represents the services in the form of URL and attributes and can work in directory-based or directory-less fashion. The work by Mallah[9] is based on a number of directory servers called virtual backbone. The author formulated a number of metrics like battery power, node's average velocity, effective degree, stability constraint and based on these parameters selected a number of virtual backbone nodes that are responsible for maintaining the directory of services.

There are a number of other directions in which research on Service Discovery in MANET has been pursued. Lenders[11] forms an analogy of MANET with electrostatic field and service as a point charge. Using this model, they formulated a service discovery algorithm and defined the required protocols and methods needed for service discovery. Torres[12] worked on service discovery at network layer and extended AODV routing protocol for service discovery in MANET. Noman[13] extended this approach further by adding a proactive component to improve latency. Hong [14] proposed service discovery based on ZeroConf. Services are advertised and browsed only in the network during topology changes or initiation of ZeroConf algorithm. Talwar[15] has proposed a service discovery algorithm based on routing intelligent mobile nodes that indexes resources and services information periodically and is queried during the discovery phase.

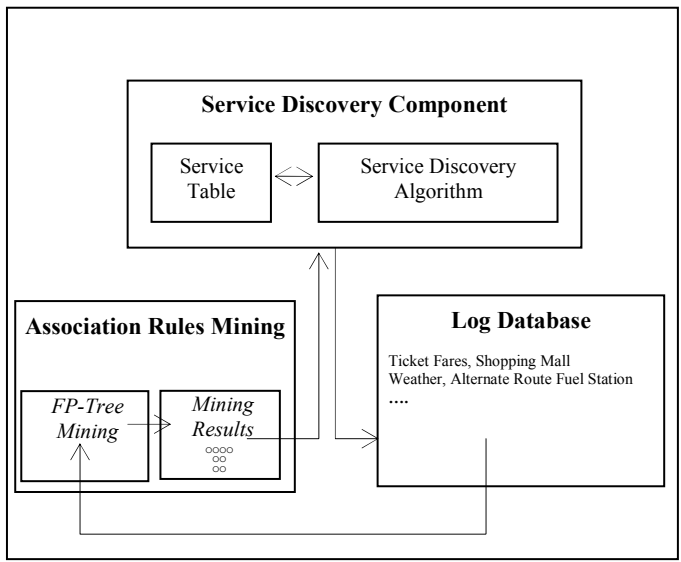

Fig 1 –Service Discovery based on Association Rules Mining

Using data mining for solving MANET issues (like service discovery) is a relatively new concept and we see only a small amount of work in this direction. Hu [16] applied the mining technique for composite web service discovery but this work was not applied to Mobile Ad hoc Network. Jabas[17] applies the mining technique to MANET traffic to find interesting relations ships among MANET nodes. Their research was motivated by the fact that these relationships can assist routing, MAC and other protocols of MANET. Our work in this paper has been done from another dimension i.e. using the association rule mining to improve the performance of Service Discovery algorithm. We use Association Rules mining technique to predict future service discovery requests based on current service discovery request. There are a two popular Association Rules mining algorithms i.e. Apriori[18] and FP-Tree[19]. Apriori algorithm is computationally too expensive. FP-Tree uses a tree like structure to reduce computational cost and is usually the recommended choice.

### III. USING ASSOCIATION RULES MINING FOR SERVICE DISCOVERY IN MANET

Fig 1 shows the proposed Service Discovery Module that is running on every node. The module comprises of a Service Discovery Component that contains a service table maintaining all the services the node knows of. The component runs a service discovery algorithm that is described in Fig 2. The details of this service

```
[Logging Component]
public void loggingComponent() {
        while(true) {
                Message m = listenForMessage();
                if(m is SREQ) {
                        Database d = openLogDatabase ();
                        if(d is full) {
                                deleteOldestRecord(d);
                        }
                        write(serviceRecord(d));
                }
        }
}

[Mining Component]
public void MiningComponent() {
        while(true) {
                readLogDatabase();
                genrateItemSetsUsingFPGrowth();
        }
}

[Service Discovery Component]
public void ServiceDiscoveryComponent() {
    while(true) {
        Message m = listenForMessage();
        if(m is SREQ) {
                if(desiredServiceAvailable) {
                        RS = getRelatedServicesIKnow();
                        generateSREPWithRS();
                }else{
                        broadCastSREQAhead();
                }
        }
        if(m is SREP) {
                if(m destination is Me) {
                        s = extractServiceInfo(m);
                        storeInServiceTable(s);
                        RS = getRelatedServiceFrom(m);
                        storeInServiceTable (RS);
                }else{
                        broadcastSREPAhead();
                        s = extractServiceInfo(m);
                        storeInServiceTable(RS);
                        RS = getRelatedServiceFrom(m);
                        storeInServiceTable(RS);
                }
        }
    }
}
```

Fig 2 – Pseudo Code describing the proposed approach

discovery algorithm are discussed in next paragraph. Every service discovery request that is generated by the service consumer will be logged in a log database. We define a service session as an interval in which a number of service discovery requests are issued by a service consumer such that these requests are related. A log database consists of a collection of session records. Each session record comprises of a number of fields and each field represents a single service discovery request issued by consumer. The Association Rules mining component uses the records stored in log database and applies FP-Tree mining algorithm on the these records to determine the correlation among the service requests. The algorithm determines what service requests are issued together i.e. frequent item sets. We have not utilized association rules since we assumed that service requests in a session can be issued in different order. The frequent item sets thus discovered can be used to predict future service discovery requests based on the current service discovery requests. Fig 2 describes the proposed approach in pseudo code format running on every node. The logging component listens for all incoming service discovery requests and records this in log database. The log database is a circular list of records. So, when the log database becomes full, it deletes starting records to log new service sessions. These logged items will then be mined using FP-Tree algorithm. The mining component reads the log database and using this information, it determines the correlation among the service discovery requests. It will determine what service requests are issued together. If the number of transactions in which some services are issued together crosses a threshold value (i.e. support), these services forms frequent item set and can be used for predicting future service requests based on current service request.

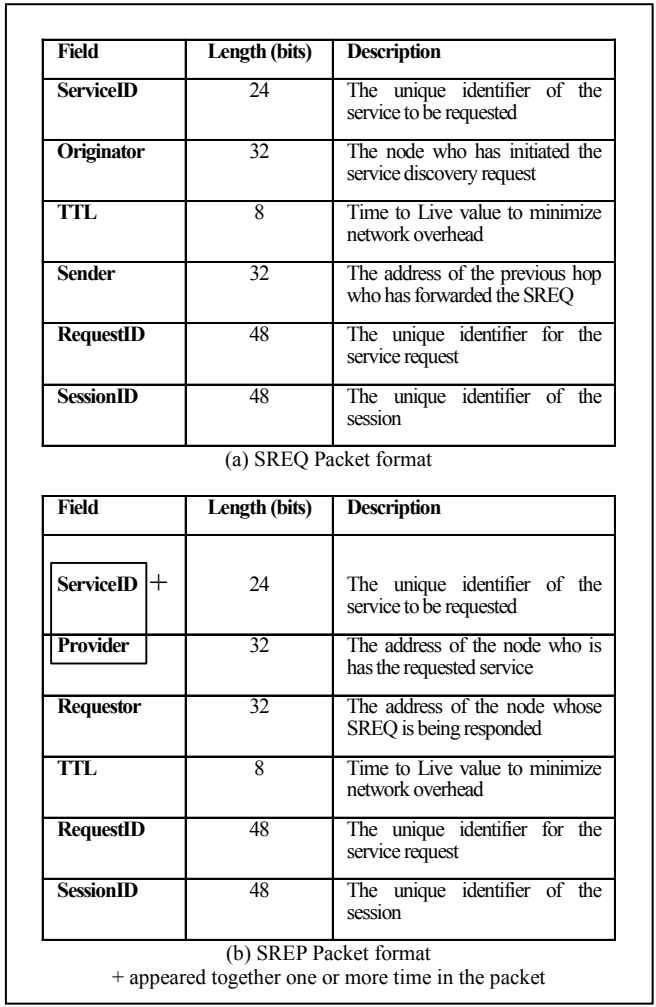

Fig 3 – Format of SREQ and SREP packets

In our approach, we have applied the mining technique on broadcast approach to service discovery. The service discovery component listens for incoming message. If the incoming message M is a service discovery request i.e. SREQ, the method extracts the requested service from the packet and looks for this service in its local service table. If the service table contains the desired service, then all the services related to this service are also picked (from the mined results) and they are also returned along with the service in SREP message. If no service is found in service table, then SREQ is propagated ahead. If the incoming message is SREP, the node sees if this is the reply of its own SREQ, then it inserts this information in its service table along with other related service information. If SREP doesn't belong to node, it propagates the SREP message along with recording service and related information in its local service table.

## IV. SIMULATION DETAILS AND RESULTS

The proposed approach has been implemented in JIST/SWANS network simulator[20, 21]. The simulation has been done in a field of $500 \times 500$ with random placement of nodes. Unless otherwise mentioned, we have used the default configuration parameters of JIST/SWANS for simulation. In order to test our algorithm, we need to generate service discovery requests in a session such that they are correlated. For that purpose, we developed a Correlated Data Generator (CDG) module that issues service discovery requests based on a correlation matrix **CM** of $n \times n$, where **n** is the number of services in the network. The correlation matrix is built programmatically using Java Random Number Generator that follows a Uniform Distribution. An $n \times n$ matrix of random numbers **R** is generated and based on this matrix, the correlation matrix **CM**, is built in the following way:

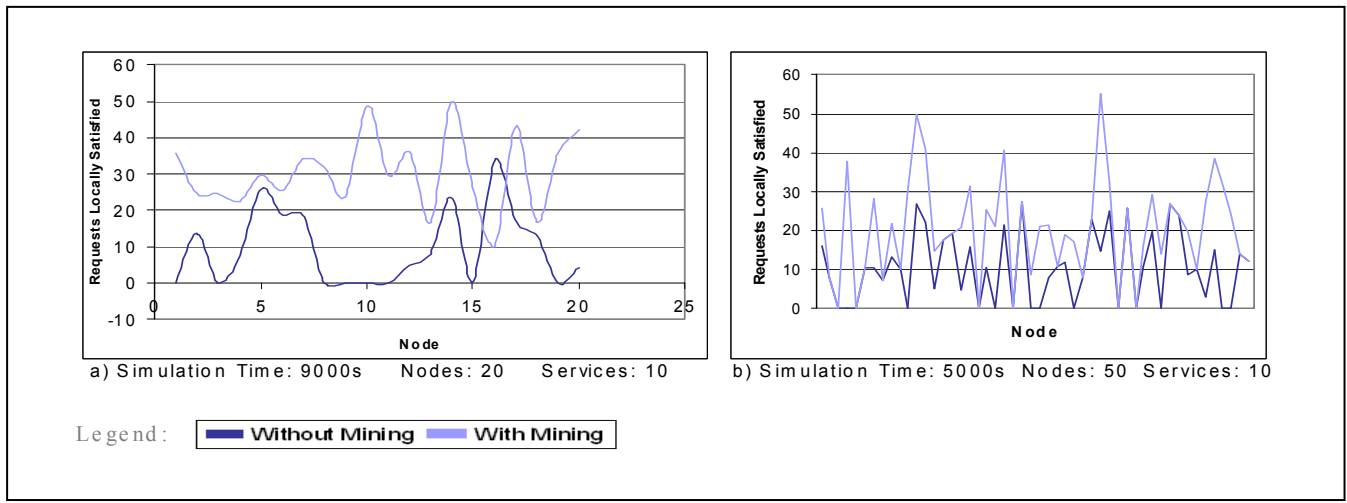

Fig 4: Results of proposed approach with AODV as routing protocol

$$CM(i,j) = \begin{cases} 0 & R(i,j) < 0.5 \\ 1 & R(i,j) \geq 0.5 \end{cases}$$

where, $0 \leq i < n$ and $0 \leq j < n$ and $0 < R(i,j) \leq 1$

Using this correlation matrix **CM**, CDG decides what service discovery requests are to be issued in one session. First, we pick a service **s** randomly. This service will be part of current session. For the sake of simplicity, by a service **s** means the index/position of the service in a sorted list of all the services available in the network. After picking a random service **s**, we select all the services related to **s** based on **CM**. These services are possible candidates for inclusion in current session. We define Candidate Service Requests set **C** as:

$$C = \{ i \mid i = s \vee CM(i,s) = 1 \}$$

To generate a session, we pick services from candidate service request and based on certain threshold, they are probabilistically issued on the network. We generate a random number $0 \leq p_i < 1$, $\forall i \in C$. A service session **S** is then defined as:

$$S = \bigcup_{p_i < \eta} C(i)$$

where, $\eta$ is the threshold value that determines the likelihood of a candidate service request to be issued to the network. We use $\eta = 0.8$. The support value for FP–Tree mining in our experiments was set to 80%. To test our approach, we have simulated the proposed scheme in an Ad hoc network with various configurations. The size of the service table or service cache was kept to 5 entries with **FIFO** strategy as the replacement algorithm. The routing algorithm used is Ad hoc On-Demand Distance Vector Routing. Fig 4a shows the simulated results with a network of 20 nodes with 10 services. The parameter we measured is the number of service discovery requests that are locally satisfied. As we can see, when we utilize data mining information, the number of service discovery requests locally satisfied increased. Fig 4b shows the results with the number of nodes increased to 50. In both of the cases, there are some minor instances where results without mining gives better results. This is because the prediction of service requests failed i.e. a node predicts a service discovery request which is not actually used in future. This caused network and processing overhead and results in degradation of performance

## V. CONCLUSION

We have proposed a novel approach to improve service discovery in MANET using Association Rules mining. The proposed approach has been applied to broadcast approach and found to give significant improvement in results. The future work in this direction is the testing and extension of the proposed approach for other parameters (e.g. latency, bandwidth) and large scale MANET. In the current approach, we have not considered the computational cost of FP-Tree mining for MANET. Future work in this direction is the development of approximate FP-Tree mining algorithm for low-powered devices. Another extension is the application of mining approach to other Service Discovery algorithms.

ACKNOWLEDGMENT

This research work is supported by 'Center of Research in Ubiquitous Computing', National University of Computer and Emerging Sciences, Karachi, Pakistan and 'Higher Education Commission', Pakistan.